\documentclass[prd,reprint,nofootinbib,superscriptaddress,preprintnumbers,showpacs]{revtex4-1}
\usepackage{amsfonts}
\usepackage{amssymb}
\usepackage{amsmath}
\usepackage{graphicx,color}
\usepackage{dcolumn}
\usepackage{epsfig}
\usepackage{bm}
\usepackage{bbm}
\usepackage{ulem}
\usepackage{hyperref}
\usepackage{lineno}

\usepackage{color}

\def\gtap{\ \raise.3ex\hbox{$>$\kern-.75em\lower1ex\hbox{$\sim$}}\ }
\def\ltap{\ \raise.3ex\hbox{$<$\kern-.75em\lower1ex\hbox{$\sim$}}\ }
\begin{document}

\title{
Pole determination of 
$P_{\psi s}^\Lambda(4338)$
and possible
$P_{\psi s}^\Lambda(4255)$
in
$B^-\to J/\psi\Lambda\bar{p}$ 
}
\author{S.X. Nakamura}
\email{satoshi@ustc.edu.cn}
\affiliation{
University of Science and Technology of China, Hefei 230026, 
People's Republic of China
}
\affiliation{
State Key Laboratory of Particle Detection and Electronics (IHEP-USTC), Hefei 230036, People's Republic of China}
\author{J.-J. Wu}
\email{wujiajun@ucas.ac.cn}
\affiliation{
School of Physical Sciences, University of Chinese Academy of Sciences (UCAS), Beijing 100049, China
}

\begin{abstract}
First hidden-charm pentaquark candidate with strangeness, $P_{\psi s}^\Lambda(4338)$, was recently discovered in $B^-\to J/\psi\Lambda\bar{p}$ by the LHCb Collaboration. 
$P_{\psi s}^\Lambda(4338)$ shows up as a bump at the $\Xi_c\bar{D}$ threshold in the $J/\psi\Lambda$ invariant mass ($M_{J/\psi\Lambda}$) distribution.
The $M_{J/\psi\Lambda}$ distribution also shows a large fluctuation at the $\Lambda_c\bar{D}_s$ threshold, hinting the existence of a possible $P_{\psi s}^\Lambda(4255)$.
In this work, we determine the $P_{\psi s}^\Lambda(4338)$ and $P_{\psi s}^\Lambda(4255)$
pole positions for the first time. 
For this purpose, we fit a $B^-\to J/\psi\Lambda\bar{p}$ model to the $M_{J/\psi\Lambda}$, $M_{J/\psi\bar{p}}$, $M_{\Lambda\bar{p}}$, and $\cos\theta_{K^*}$ distributions from the LHCb simultaneously; $\chi^2/{\rm ndf}\sim 1.21$.
Then we extract $P_{\psi s}^\Lambda$ poles from a unitary $\Xi_c\bar{D}$-$\Lambda_c\bar{D}_s$ coupled-channel scattering amplitude built in the model.
In our default fit, 
the $P_{\psi s}^\Lambda(4338)$ pole is found at $( 4338.2\pm 1.4)-( 1.9\pm 0.5 )\,i$~MeV while 
the $P_{\psi s}^\Lambda(4255)$ pole at $4254.7\pm 0.4$~MeV.
The $P_{\psi s}^\Lambda(4338)$ and $P_{\psi s}^\Lambda(4255)$ are mostly 
$\Xi_c \bar{D}$ bound and 
$\Lambda_c\bar{D}_s$ virtual states, respectively.
Through our analysis, the data disfavors a hypothesis of $P_{\psi s}^\Lambda(4338)$ as merely a kinematical effect.
This pole determination, which is important in its own right, sets a primary basis to study the nature of the $P_{\psi s}^\Lambda$ states.
\end{abstract}

\maketitle

\section{Introduction}
Since the foundation of the quark model, we have been addressing the fundamental question:
``What form of the matter can be built from quarks~?''
Recent experimental discoveries of pentaquark and tetraquark candidates have decisively widened our territory of the conventional $qqq$ and $q\bar{q}$ hadrons to include qualitatively different $qqqq\bar{q}$, $qq\bar{q}\bar{q}$, and even more exotic structures;
see reviews~\cite{review_chen,review_hosaka,review_lebed,review_esposito,review_ali,review_guo,review_olsen,review_Brambilla}.
Establishing the (non-)existence of pentaquark and tetraquark states is now essential to answer the above fundamental question.

The existence of hidden-charm pentaquarks with strangeness ($P_{\psi s}^\Lambda$, $udsc\bar{c}$) has been expected theoretically~\cite{Wu:2010jy, Wu:2010vk, Yuan:2012wz}, and the discovery of hidden-charm pentaquark candidates
($uudc\bar{c}$)~\cite{lhcb_pc1,lhcb_pc2} further strengthened the expectation~\cite{Xiao:2019gjd, Chen:2015sxa, Santopinto:2016pkp, Shen:2019evi, Chen:2016ryt, Wang:2019nvm}.
The first evidence (3.1$\sigma$) of $P_{\psi s}^\Lambda$ was found, by the LHCb Collaboration, in $\Xi^-_b \to J/\psi\Lambda K^-$ as a bump at $\sim 4459$~MeV in the $J/\psi\Lambda$ invariant mass ($M_{J/\psi\Lambda}$) distribution~\cite{lhcb_pcs4459}.
This result invited lots of theoretical studies on $P_{\psi s}^\Lambda(4459)$~\cite{Xiao:2021rgp, Peng:2020hql, Clymton:2021thh, Wang:2021itn, Lu:2021irg, Li:2021ryu, Hu:2021nvs, Chen:2021cfl, Cheng:2021gca, Chen:2020uif, Wang:2020eep, Azizi:2021utt, Peng:2020hql, Chen:2020kco, Liu:2020hcv, Liu:2020ajv}.
Then, very recently, the LHCb announced the first discovery ($>10\sigma$) of $P_{\psi s}^\Lambda$ in $B^-\to J/\psi\Lambda\bar{p}$~\cite{lhcb_seminar}.
Their amplitude analysis determined the $P_{\psi s}^\Lambda$ mass,
width, and spin-parity 
 as $ 4338.2\pm 0.7~{\rm MeV}$, $ 7.0\pm 1.2~{\rm MeV}$, and $J^P=1/2^-$,
respectively.
%
In response to the discovery, proposals have been made to interpret $P_{\psi s}^\Lambda(4338)$
as a $\Xi_c\bar{D}$ molecule~\cite{Karliner2022,fwang2022,myan2022}, and as a triangle singularity~\cite{pcs4338_ts}.

To understand the nature of $P_{\psi s}^\Lambda(4338)$, its properties such as the mass, width, and $J^P$ are the crucial information.
The LHCb amplitude analysis obtained them under an assumption that the $P_{\psi s}^\Lambda(4338)$ peak is due to a resonance that can be well simulated by a Breit-Wigner (BW) amplitude.
However, the $P_{\psi s}^\Lambda(4338)$ peak is located right on the $\Xi_c\bar{D}$ threshold [see Fig.~\ref{fig:comp-data}(a)], which would invalidate this assumption.
First of all, the resonancelike structure might be caused by a kinematical effect (threshold cusp) and not by a resonance pole~\cite{ts_review}.
If a $P_{\psi s}^\Lambda(4338)$ pole exists and it couples with $\Xi_c\bar{D}$, the BW amplitude is still not suitable since it does not account for:
(i) the width starts to increase rapidly as the $\Xi_c\bar{D}$ channel opens;
(ii) the lineshape due to the pole can be distorted by the branch point (threshold)
of the complex energy plane where the pole is located.
The relevance of the items (i,ii) was demonstrated in Ref.~\cite{lmeng2022}.

What needs to be done is to replace the BW approximation with the proper pole-extraction method where a unitary coupled-channel amplitude is fitted to the data, and poles on relevant Riemann sheets are searched by analytically continuing the amplitude.
This is the main task in this Letter.
The pole value not only provides an important knowledge reflecting the QCD dynamics but also serves as a basis for studying the nature of $P_{\psi s}^\Lambda(4338)$.

Meanwhile, the $M_{J/\psi\Lambda}$ distribution data shows a
large fluctuation at $M_{J/\psi\Lambda}\sim 4255$~MeV.
The LHCb examined this fluctuation as a possible 
$P_{\psi s}^\Lambda(4255)$ contribution, and found it statistically insignificant.
However, the fluctuation occurs at 
the $\Lambda_c \bar{D}_s$ threshold where
a visible threshold cusp would be expected
from a color-favored $B^-\to \Lambda_c \bar{D}_s \bar{p}$ followed by 
 $\Lambda_c \bar{D}_s\to J/\psi \Lambda$.
A $\Lambda_c \bar{D}_s$ rescattering might cause a pole to 
enhance the cusp.

\begin{figure*}[t]
\begin{center}
\includegraphics[width=1\textwidth]{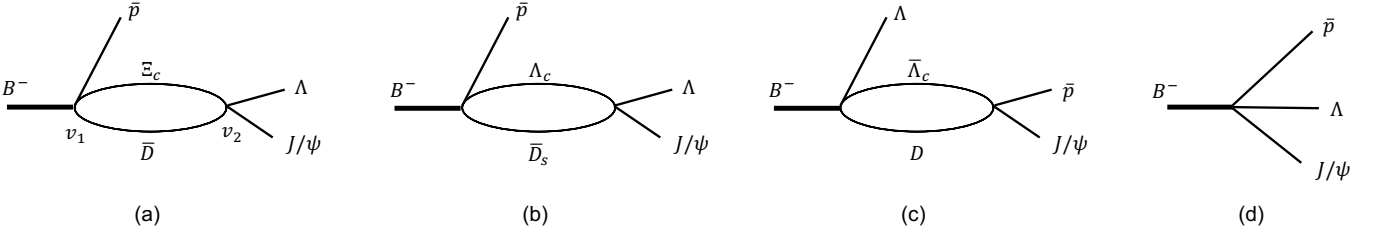}
\end{center}
 \caption{
$B^-\to J/\psi \Lambda \bar{p}$ mechanisms 
initiated by weak vertex $v_1$ of 
(a) $B^-\to\Xi_c\bar{D}\bar{p}$;
(b) $B^-\to\Lambda_c\bar{D}_s\bar{p}$;
(c) $B^-\to\bar{\Lambda}_c D\Lambda$;
(d) $B^-\to J/\psi\Lambda\bar{p}$.
The second vertex $v_2$ in (a,b) [(c)] includes a
 $\Xi_c\bar{D}-\Lambda_c\bar{D}_s$ coupled-channel 
 [$\bar{\Lambda}_c D$ single-channel] scattering, followed by a perturbative
 transition to $J/\psi\Lambda$ [$J/\psi\bar{p}$].
 }
\label{fig:diag}
\end{figure*}

In this work, we analyze the LHCb data on $B^-\to J/\psi\Lambda\bar{p}$
in detail.
The $M_{J/\psi\Lambda}$, $M_{J/\psi\bar{p}}$, $M_{\Lambda\bar{p}}$, and $\cos\theta_{K^*}$
distribution data are simultaneously fitted with a model in which a unitary $\Xi_c\bar{D}$-$\Lambda_c\bar{D}_s$ coupled-channel amplitude is implemented. 
Based on the coupled-channel amplitude, we address the following issues:
(i) the $P_{\psi s}^\Lambda(4338)$ pole position;
(ii) a possibility that the $P_{\psi s}^\Lambda(4338)$ peak is 
merely a $\Xi_c \bar{D}$ threshold cusp;
(iii) implications of the large fluctuation at the $\Lambda_c\bar{D}_s$ threshold.
\section{Model}
The LHCb data shows visible structures only around the $\Xi_c\bar{D}$, $\Lambda_c\bar{D}_s$, and 
$\bar{\Lambda}_c D$ thresholds.
Thus it is reasonable to
assume that the structures are caused by the threshold cusps that are further enhanced or suppressed by hadronic rescatterings and the associated poles~\cite{xkdong};
see Figs.~\ref{fig:diag}(a-c).
Other possible mechanisms are assumed to be absorbed by a direct decay
mechanism of Fig.~\ref{fig:diag}(d).
For the small $Q$-value ($\sim 130$~MeV) in
$B^-\to J/\psi\Lambda\bar{p}$,
we consider only $s$-wave interactions that are expected to
dominate. 
We confirmed that $p$-wave spectators in 
Fig.~\ref{fig:diag} hardly improve fitting the data.

We present amplitude formulas for Figs.~\ref{fig:diag}(a-c).
The energy, momentum, and polarization vector of a particle $x$ are denoted by $E_x$, $\bm{p}_x$, and $\bm{\epsilon}_x$, respectively, and particle masses are from Ref.~\cite{pdg}.
We also denote a baryon($B$)-meson($M$) pair with $J^P$ by $BM(J^P)$.
The initial weak $B^-\to \Xi_c\bar{D}(1/2^-)\bar{p}$ vertex [Fig.~\ref{fig:diag}(a)] is
\begin{eqnarray}
\label{eq:dt11}
v_{1}&=& c^{1/2^-}_{\Xi_c \bar{D}\bar{p},B^-}\,
\langle t_{\bar{D}}  t^z_{\bar{D}} t_{\Xi_c} t^z_{\Xi_c} | 00\rangle
 f_{\Xi_c \bar{D}}^{0}
 F_{\bar{p}B^-}^{0}\ , 
\end{eqnarray}
with a complex coupling constant
$c^{1/2^-}_{\Xi_c \bar{D}\bar{p},B^-}$.
An isospin Clebsch-Gordan coefficient is given by the bracket where $t^{(z)}_x$ is the isospin ($z$-component) of a particle $x$.
The $B^-\to \Lambda_c\bar{D}_s(1/2^-)\bar{p}$ [Fig.~\ref{fig:diag}(b)] and $\bar{\Lambda}_c D^0 (1/2^+)\Lambda$ [Fig.~\ref{fig:diag}(c)] vertices are the same form with couplings 
$c^{1/2^-}_{\Lambda_c \bar{D}_s\bar{p},B^-}$ and $c^{1/2^+}_{\bar{\Lambda}_c D   \Lambda,B^-}$.
We introduced dipole form factors $f_{ij}^{L}$ and $F_{kl}^{L}$ defined by
\begin{eqnarray}
\label{eq:ff1}
 f_{ij}^{L} &=&
 {(1+q_{ij}^2/\Lambda^2)^{-2-{L\over 2}}\over \sqrt{E_i E_j}} ,
F_{kl}^{L} =
 {(1+\tilde{p}_k^2/\Lambda^{2})^{-2-{L\over 2}}\over \sqrt{E_k E_l}} ,
\end{eqnarray}
where $q_{ij}$ ($\tilde{p}_{k}$) is the momentum of $i$ ($k$) in the $ij$ (total) center-of-mass frame.
We use a common cutoff value $\Lambda=1$~GeV in Eq.~(\ref{eq:ff1}) for all the interaction vertices.
The $B^-\to\bar{\Lambda}_c D^0\Lambda, \Lambda_c\bar{D}_s\bar{p}$ decays are color-favored processes, while $B^-\to \Xi_c\bar{D}\bar{p}$ is color-suppressed.

The above weak decays are followed by hadronic scatterings.
We take a data-driven approach to the hadron interactions while respecting the relevant coupled-channel unitarity; the idea on which the $K$-matrix approach is also based.
We use hadron interactions in a form not biased by any particular models, and all coupling strengths are determined by the data. 

We consider the most important coupled-channels:
a $\Xi_c\bar{D}-\Lambda_c\bar{D}_s(1/2^-)$ coupled-channel scattering in Figs.~\ref{fig:diag}(a,b), and a $\bar{\Lambda}_c D(1/2^+)$ single-channel scattering in Fig.~\ref{fig:diag}(c).
We assume that transitions to the $J/\psi\Lambda$ and $J/\psi\bar{p}$ channels can be treated perturbatively.
We use an $s$-wave meson-baryon interaction potential:
\begin{eqnarray}
v_{\beta,\alpha} &=& h_{\beta,\alpha}
\langle t_{\beta 1} t^z_{\beta 1} t_{\beta 2} t^z_{\beta 2} | TT^z \rangle
\langle t_{\alpha 1} t^z_{\alpha 1} t_{\alpha 2} t^z_{\alpha 2} | TT^z\rangle
\nonumber \\
&&\times f^0_\beta Y_{00}\,  f^0_\alpha Y_{00},
\, 
\label{eq:cont-ptl}
\end{eqnarray}
where $\alpha$ and $\beta$ label coupled-channels such as $\Xi_c\bar{D}(1/2^-)$, and ${\alpha 1}$ and ${\alpha 2}$ are the meson and baryon in a channel $\alpha$, respectively;
$h_{\beta,\alpha}$ is a coupling constant;
$Y_{lm}$ denotes a spherical harmonics. 
We introduce $[G^{-1}(E)]_{\beta\alpha}=\delta_{\beta\alpha} - h_{\beta,\alpha} \sigma_\alpha(E)$ with
\begin{eqnarray}
\sigma_\alpha(E) &=&
\sum_{t^z}  \int\! dq q^2 
{ \langle t_{\alpha 1} t^z_{\alpha 1} t_{\alpha 2} t^z_{\alpha 2} |TT^z\rangle^2
 \left[f^0_\alpha(q)\right]^2 
\over E-E_{\alpha 1}(q)-E_{\alpha 2}(q) 
+ i\varepsilon} ,
\label{eq:sigma}
\end{eqnarray}
where $\sum_{t^z}$ is needed for $\alpha=\Xi_c \bar{D}$; $\Xi_c^{+} D^-$ and $\Xi_c^0 \bar{D}^0$ intermediate states with the charge dependent masses are included.
The perturbative interactions for $\Xi_c\bar{D}(1/2^-), \Lambda_c\bar{D}_s(1/2^-)\to J/\psi \Lambda$ and $\bar{\Lambda}_c D^0 (1/2^+)\to J/\psi \bar{p}$ are given by $s$-wave separable interactions:
\begin{eqnarray}
\label{eq:v2}
v_{\gamma,\alpha} &=& 
h_{\gamma, \alpha}
\langle t_{\alpha 1} t^z_{\alpha 1} t_{\alpha 2} t^z_{\alpha 2} |TT^z\rangle\,
\bm{\sigma}\cdot \bm{\epsilon}_\psi \,
 f_{\gamma}^{0}Y_{00}\, f_{\alpha}^{0}Y_{00}\ , 
\end{eqnarray}
where $\gamma=J/\psi \Lambda$ or $J/\psi \bar{p}$, and
$\bm{\sigma}$ is the Pauli matrix.

With the above ingredients,
following the time-ordered perturbation theory,
the amplitudes are given as
\begin{widetext}
\begin{eqnarray}
A^{\rm loop}_{\psi\Lambda(1/2^-)} &=&
\sum_{\alpha,\beta}^{\Xi_c\bar{D},\Lambda_c\bar{D}_s}
h_{\psi \Lambda, \beta}\,
c^{1/2^-}_{\alpha\bar{p},B^-}\,
\bm{\sigma}\cdot \bm{\epsilon}_\psi \,
 f_{\psi \Lambda}^{0}(p_\psi)\,
\sigma_{\beta}(M_{\psi\Lambda})\,
G_{\beta\alpha}(M_{\psi\Lambda})
 F_{\bar{p}B^-}^{0}
\ ,
\label{eq:1L1}
\end{eqnarray}
for Figs.~\ref{fig:diag}(a) and \ref{fig:diag}(b), and
\begin{eqnarray}
%
A^{\rm loop}_{\psi\bar{p}(1/2^+)} &=&
h_{\psi \bar{p}, \bar{\Lambda}_c D}\,
c^{1/2^+}_{\bar{\Lambda}_c D\Lambda,B^-}\,
\bm{\sigma}\cdot \bm{\epsilon}_\psi \,
 f_{\psi \bar{p}}^{0}(p_\psi)\,
\sigma_{\bar{\Lambda}_c D}(M_{\psi\bar{p}})\,
G_{\bar{\Lambda}_c D,\bar{\Lambda}_c D}(M_{\psi\bar{p}})
 F_{\Lambda B^-}^{0}
\ ,
\label{eq:1L2}
\end{eqnarray}
\end{widetext}
for Fig.~\ref{fig:diag}(c).
The spinors of the final $\Lambda$ and $\bar{p}$ implicitly sandwich the above expressions.

Regarding the color-suppressed direct decay mechanism of Fig.~\ref{fig:diag}(d),
we find the following $J/\psi\bar{p}(1/2^+)$ partial wave amplitude
gives a reasonable fit:
\begin{eqnarray}
\label{eq:dir_s}
 A^{\rm dir}_{\psi\bar{p}(1/2^+)} &=&
  c^{1/2^+}_{\psi\bar{p} \Lambda,B^-}
\bm{\sigma}\cdot \bm{\epsilon}_\psi  \,
 f_{\psi\bar{p}}^{0} 
 F_{\Lambda B^-}^{0} \ , 
\end{eqnarray}
with a coupling constant $c^{1/2^+}_{\psi\bar{p}\Lambda,B^-}$.

\section{Results}
We use the amplitudes of Eqs.~(\ref{eq:1L1})--(\ref{eq:dir_s}) to
 simultaneously fit the $M_{J/\psi\Lambda}$, $M_{J/\psi\bar{p}}$, $M_{\Lambda\bar{p}}$, and
$\cos\theta_{K^*}$ distributions~\footnote{
$\cos\theta_{K^*}\equiv {\bm{p}_\Lambda\cdot \bm{p}_\psi\over |\bm{p}_\Lambda||\bm{p}_\psi|}$ 
in the $\Lambda\bar{p}$ center-of-mass frame.
} from the LHCb;
see Appendix~B of Ref.~\cite{3pi}
for the procedure of calculating the invariant mass distributions.
Theoretical invariant mass ($\cos\theta_{K^*}$)
distributions are smeared with experimental resolutions of 1~MeV 
(bin width of 0.05), and are further averaged over the bin width in each
bin. 
The obtained binned theoretical distributions are used to 
calculate $\chi^2$.
The amplitudes include adjustable coupling constants from the weak
vertices of Eqs.~(\ref{eq:dt11}) and (\ref{eq:dir_s}),
and from hadronic interactions of
Eqs.~(\ref{eq:cont-ptl}) and (\ref{eq:v2}).
We reduce the fitting parameters by
setting $h_{\psi \Lambda, \Xi_c\bar{D}}=h_{\psi\Lambda, \Lambda_c\bar{D}_s}$
since the fit quality does not significantly change by allowing 
$h_{\psi \Lambda, \Xi_c\bar{D}}\ne h_{\psi\Lambda,\Lambda_c\bar{D}_s}$.
Then we adjust the products such as 
$h_{\psi \Lambda, \alpha}\,c^{1/2^-}_{\Xi_c\bar{D}\bar{p},B^-}$
in Eqs.~(\ref{eq:1L1}) and (\ref{eq:1L2}).
Numerical values for the coupling constants determined by the fit 
are given in the Supplemental Material.
Our default model has 9 fitting parameters in total, considering that the magnitude
and phase of the full amplitude are arbitrary.

\begin{figure}[b]
\begin{center}
\includegraphics[width=.5\textwidth]{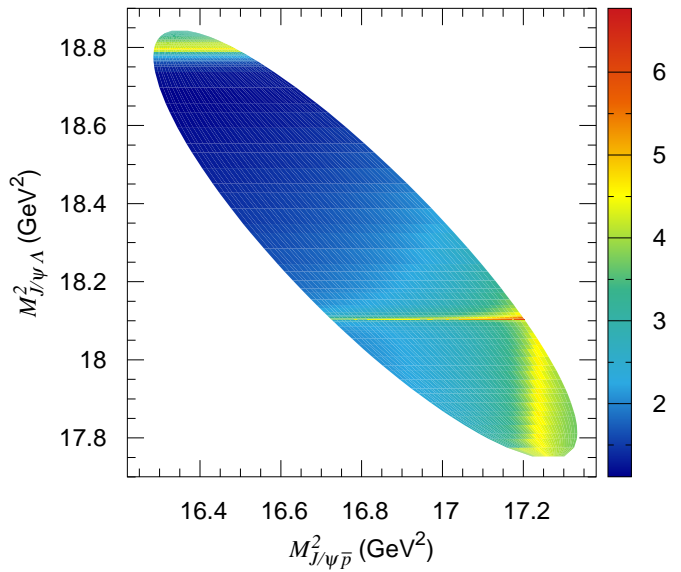}
\end{center}
 \caption{
$B^-\to J/\psi\Lambda\bar{p}$ 
Dalitz plot distribution from the default model. No smearing is applied.
 }
\label{fig:dalitz}
\end{figure}
We first present a Dalitz plot distribution
from the default model in Fig.~\ref{fig:dalitz}.
Comparing the plot with the LHCb's (Fig.~2 in Ref.~\cite{lhcb_seminar}),
the overall pattern is quite similar, except that the peak structures in
our plot are sharper since no smearing with the experimental resolution
is considered.

\begin{figure*}[t]
\begin{center}
\includegraphics[width=1\textwidth]{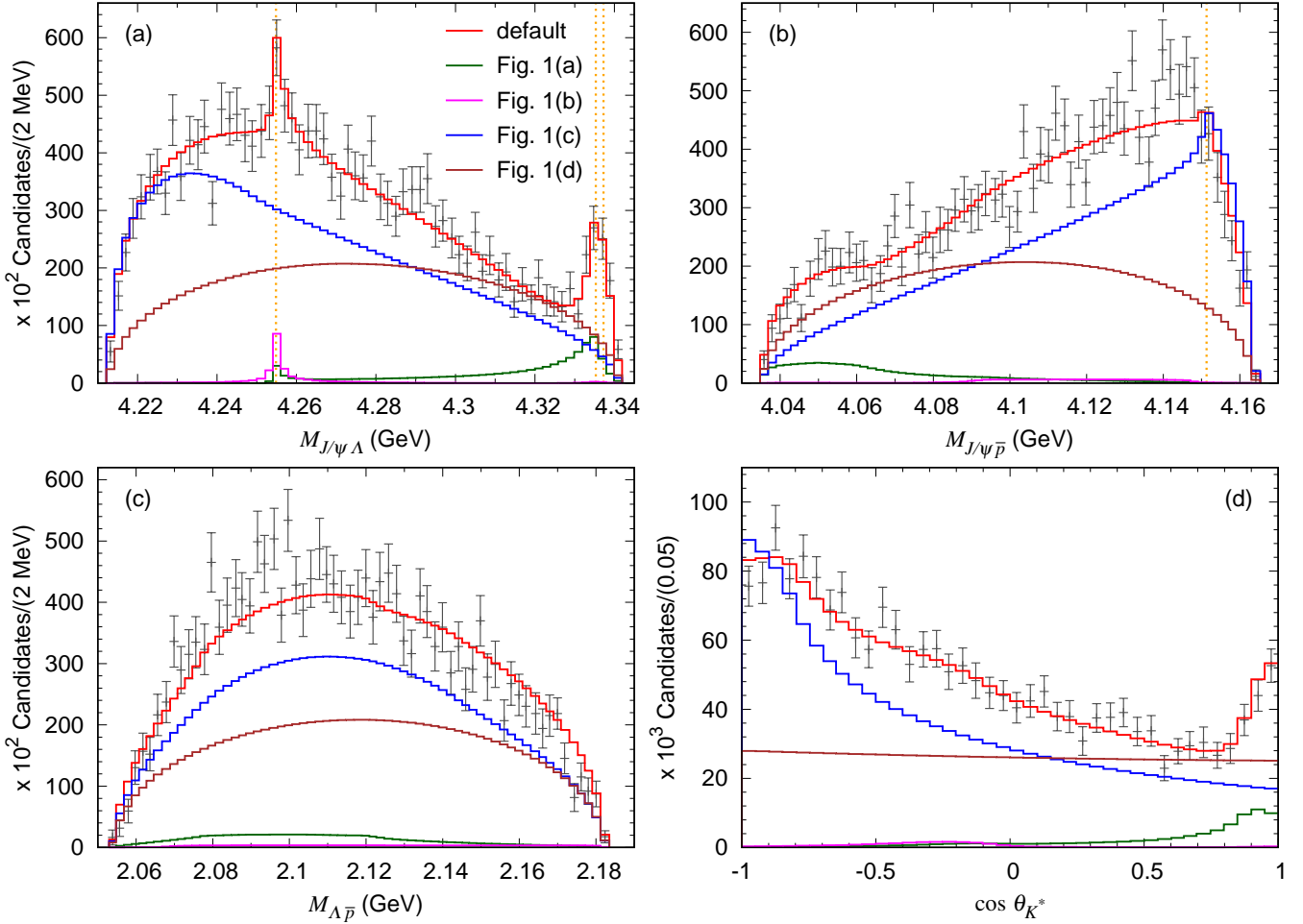}
\end{center}
 \caption{
(a) $J/\psi \Lambda$, (b) $J/\psi \bar{p}$,
(c) $\Lambda \bar{p}$ invariant mass,
and (d) $\cos\theta_{K^*}$ distributions
  for $B^-\to J/\psi \Lambda \bar{p}$.
The default fit and 
contributions from different diagrams in Fig.~\ref{fig:diag}
are shown.
The dotted vertical lines indicate thresholds for,
 from left to right, 
$\Lambda_c^+ D_s^-$, 
$\Xi_c^0 \bar{D}^{0}$, and
$\Xi_c^+ D^-$ 
[$\bar{\Lambda}_c^- D^0$]
in the panel (a) [(b)].
Data~\cite{lhcb_seminar} are
efficiency corrected and background subtracted.
 }
\label{fig:comp-data}
\end{figure*}

Now, we compare the default model (red histogram) with the LHCb
data in Fig.~\ref{fig:comp-data}, showing a good agreement.
In particular, the $P_{\psi s}^\Lambda(4338)$ and possible $P_{\psi s}^\Lambda(4255)$ peaks 
in the $M_{J/\psi\Lambda}$ distribution are well-fitted.
The fit quality is $\chi^2/{\rm ndf}=(50+81+112+29)/(235-9)\simeq 1.21$ where four $\chi^2$s are from comparing with the $M_{J/\psi\Lambda}$, $M_{J/\psi\bar{p}}$, $M_{\Lambda\bar{p}}$, and
$\cos\theta_{K^*}$ distributions, respectively;
'ndf' is the number of bins (40 for $\cos\theta_{K^*}$ and $3\times 65$ for the others) subtracted by the number of the fitting parameters.

We also show contributions from the diagrams of Fig.~\ref{fig:diag} each
of which has a different initial weak vertex.
Overall, the diagrams of Figs.~\ref{fig:diag}(c) [blue] and \ref{fig:diag}(d) [brown] dominate the process.
The increasing $M_{J/\psi\bar{p}}$ distribution
in Fig.~\ref{fig:comp-data}(b) is understood as the $\bar{\Lambda}_c D$ threshold cusp from Fig.~\ref{fig:diag}(c)~\footnote{
The fit favors a repulsive $\bar{\Lambda}_c D$ interaction, consistent with our previous finding from analyzing $B^0_s\to J/\psi p\bar{p}$~\cite{sxn_Bs}.}.
Although the diagrams of Figs.~\ref{fig:diag}(a) [green] and
\ref{fig:diag}(b) [magenta] are relatively small in the magnitude, they
have significantly enhanced $\Xi_c\bar{D}$ and $\Lambda_c\bar{D}_s$
threshold cusps, respectively, and develop the  
$P_{\psi s}^\Lambda$ peaks through the interference. 

The large contribution from Fig.~\ref{fig:diag}(c) is understandable
since it is color favored.
The color-suppressed Fig.~\ref{fig:diag}(d) is comparable, possibly
because it is not suppressed by a loop.
However, the color-favored Fig.~\ref{fig:diag}(b) contributes rather small.
This might be because $\Lambda_c\bar{D}_s \to \Lambda J/\psi$ is suppressed compared with $\bar{\Lambda}_c D \to \bar{p} J/\psi$.
The suppression would be expected since, in a meson-exchange picture, $\Lambda_c\bar{D}_s \to \Lambda J/\psi$ caused by a $D_s^{(*)}$-exchange involves $s\bar{s}$ creation and annihilation while $\bar{\Lambda}_c D \to \bar{p} J/\psi$ with a $D^{(*)}$-exchange needs light quark pair (de)excitations.
Yet, a solid understanding awaits more theoretical analyses and higher statistics data.

Our default and LHCb's models describe the data rather differently. 
The LHCb fitted the $M_{J/\psi\bar{p}}$ distribution
with a non-resonant $J/\psi\bar{p}$ [NR($J/\psi\bar{p}$)] amplitude in a
polynomial form without identifying the physical origin of the increasing
behavior. 
The NR($J/\psi\bar{p}$) includes only a $S=3/2$ and $p$-wave
$J/\psi\bar{p}$ to occupy $\sim 84$\% fit fraction;
$S$ denotes the total spin of $J/\psi\bar{p}$.
This $p$-wave dominance is counterintuitive since 
the small $Q$-value implies a $s$-wave dominance. 
Indeed, Figs.~\ref{fig:diag}(c,d), which is dominant in our model, are
$s$-wave $J/\psi\bar{p}$ amplitudes.

We modified our default model by replacing 
Figs.~\ref{fig:diag}(c,d) with diagrammatically similar ones that have
the same $J/\psi \bar{p}$ partial wave as the LHCb's NR($J/\psi\bar{p}$).
By fitting this modified model to the LHCb data, we obtained $\chi^2$
similar to that of the default model. However, the modified and
LHCb models have lineshapes qualitatively different from the default
model such as: 
(i) At the $\bar{\Lambda}_c D$ threshold
in the $M_{J/\psi\bar{p}}$ distribution, our default fit has a
cusp structure but
our modified and the LHCb
fits have a smooth lineshape;
(ii) Our default model (and also the LHCb data tend to) shows
a plateau at $\cos \theta_{K^*}\sim -1$, but our modified and the LHCb models show a
monotonically decreasing behavior.
The future higher statistic data might distinguish these
differences.

Also, the LHCb's model has 16 fitting parameters,
in contrast to 9(8) parameters in our default model (alternative model below).
Although the LHCb's model fits richer six-dimensional data,
$\sim 2$ times more parameters sound too many.
The $p$-wave dominance and excessive parameters 
in the LHCb's model are possibly from
missing relevant mechanisms such as 
Figs.~\ref{fig:diag}(a-c);
many other mechanisms mimic the relevant ones
through complicated interferences.

\begin{table}[t]
\renewcommand{\arraystretch}{1.6}
\tabcolsep=2.mm
\caption{\label{tab:pole}
$P_{\psi s}^\Lambda$ poles.
(I) default model; (II) [(III)]
 alternative model with $v_{\Lambda_c\bar{D}_s,\Lambda_c\bar{D}_s}=0$
[energy dependence of
Eq.~(\ref{eq:energy-dep}) and $v_{\Lambda_c\bar{D}_s,\Lambda_c\bar{D}_s}=0$].
Pole positions (in MeV) and their Riemann sheets (see the text for
 notation) are given in the third and
 fourth columns, respectively.
}
\begin{tabular}{clcc}
(I)  & $P_{\psi s}^\Lambda(4338)$ &$( 4338.2\pm 1.4)-( 1.9\pm 0.5 )\,i$ & $(upp)$\\
     & $P_{\psi s}^\Lambda(4255)$ &$ 4254.7\pm 0.4                    $ & $(upp)$\\ \hline
(II)&  $P_{\psi s}^\Lambda(4338)$ &$( 4331.9\pm 4.2)+( 5.6\pm 6.4 )\,i$& $(ppu)$\\
& & $( 4328.6\pm 4.2)+( 4.6\pm 5.9 )\,i$& $(pup)$\\
& & $( 4336.1\pm 1.3)+( 0.3\pm 1.3 )\,i$& $(puu)$\\ \hline
(III) & $P_{\psi s}^\Lambda(4338)$ & $( 4340.0\pm 8.5)-( 2.2\pm 8.8 )\,i$& $(upp)$\\
& & $( 4340.1\pm 13.3)-( 5.1\pm 3.5 )\,i$& $(uup)$\\
& & $( 4338.0\pm 4.1)-( 6.2\pm 7.3 )\,i$& $(uuu)$
\end{tabular}
\end{table}

\begin{figure}[b]
\begin{center}
\includegraphics[width=.5\textwidth]{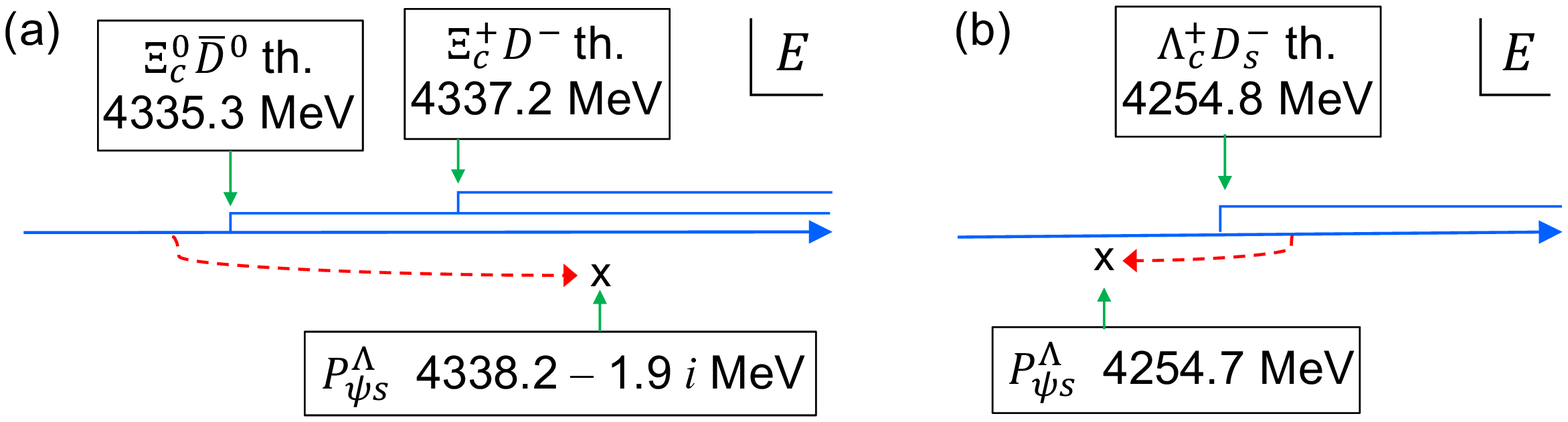}
\end{center}
 \caption{
Pole locations of (a) $P_{\psi s}^\Lambda(4338)$ and (b) $P_{\psi s}^\Lambda(4255)$ of the default model. 
The red dotted arrows indicate how to reach the poles from the closest physical energy regions.
The double lines indicate the branch cuts.
 }
 \label{fig:pole_sheet}
\end{figure}

We searched for poles in our default $\Xi_c\bar{D}-\Lambda_c\bar{D}_s(1/2^-)$ coupled-channel scattering amplitude by the analytic continuation.
We found $P_{\psi s}^\Lambda(4338)$ and $P_{\psi s}^\Lambda(4255)$ poles, as summarized in 
Table~\ref{tab:pole}; $J^P$ is consistent with the LHCb's result for $P_{\psi s}^\Lambda(4338)$.
In the table, we also list the Riemann sheets of the poles by $(s_{\Lambda_c\bar{D}_s}\,s_{\Xi^0_c\bar{D}^0}\,s_{\Xi^+_c D^-})$ where $s_\alpha=p$ or $u$ depending on whether the pole is located on the physical ($p$) or unphysical ($u$) sheet of a channel $\alpha$~\footnote{Section~50 in Ref.~\cite{pdg} defines (un)physical sheet.}.
The pole locations relative to the relevant thresholds are illustrated in Fig.~\ref{fig:pole_sheet}.
The $P_{\psi s}^\Lambda(4338)$ pole is mainly generated by
$v_{\Xi_c\bar{D},\Xi_c\bar{D}}$.
In fact, if $v_{\Xi_c\bar{D},\Lambda_c\bar{D}_s}$ is turned off, we
find a $\Xi_c \bar{D}$ bound pole at 4334.9~MeV.
On the other hand, $v_{\Lambda_c\bar{D}_s,\Lambda_c\bar{D}_s}$ alone is not strong enough to
create a $\Lambda_c\bar{D}_s$ bound state but a virtual pole at 4251.8~MeV.

A light vector-meson exchange would (not) cause a strong attraction in 
$v_{\Xi_c\bar{D},\Xi_c\bar{D}}$ ($v_{\Lambda_c\bar{D}_s,\Lambda_c\bar{D}_s}$)~\cite{Xiao:2021rgp}.
A possible mechanism to cause the relatively strong 
$v_{\Lambda_c\bar{D}_s,\Lambda_c\bar{D}_s}$ 
is a two-pion-exchange (TPE).
TPE mechanisms could be important to understand possible bound states
of a bottomonia-pair~\cite{tpe_bb} and a $J/\psi$-$J/\psi$ pair~\cite{tpe_cc}.
Also, a lattice QCD~\cite{hal_phiN} found that a TPE is the dominant long-range part of the $\phi$-nucleon interaction, causing a large attraction. 
In addition, a $K^*$-exchange $\Lambda_c\bar{D}_s \to \Xi_c\bar{D}$
provides an attraction to $v_{\Xi_c\bar{D},\Lambda_c\bar{D}_s}$.

\begin{table*}[t]
\renewcommand{\arraystretch}{1.6}
\tabcolsep=3.mm
\caption{\label{tab:para3}
Parameter values for $B^-\to J/\psi \Lambda\bar{p}$ models.
The second, third, and fourth columns are for 
the default, alternative 
($h_{\Lambda_c\bar{D}_s,\Lambda_c\bar{D}_s}=0$), 
and no-pole models, respectively.
For the arbitrariness, we may multiply a common overall complex factor
 to the parameters in the 1-4th rows.
$h_{\psi\Lambda,\alpha}=h_{\psi\Lambda,\Lambda_c\bar{D}_s}=h_{\psi\Lambda,\Xi_c\bar{D}}$.
}
\begin{tabular}{lccc}
$h_{\psi\Lambda,\alpha}\,c^{1/2^-}_{\Xi_c \bar{D}\bar{p},B^-}$        & $                  (  0.81\pm  0.23 )\,i$ &$( -1.24\pm  0.27)+(  0.40\pm  0.74 )\,i$ &$(  0.42\pm  1.24)+(  7.88\pm  0.47 )\,i$\\
$h_{\psi\Lambda,\alpha}\,c^{1/2^-}_{\Lambda_c \bar{D}_s\bar{p},B^-}$  & $( -0.24\pm  0.07)+(  0.28\pm  0.14 )\,i$ &$                  ( -1.40\pm  0.28 )\,i$ &$  -2.09\pm  0.28                       $\\
$h_{\psi\bar{p},\bar{\Lambda}_c D}\, c^{1/2^+}_{\bar{\Lambda}_c D   \Lambda,B^-}$ & $( -1.28\pm  4.23)+( 11.48\pm  3.00 )\,i$ &$( -6.19\pm  3.72)+( 10.32\pm  2.69 )\,i$ &$(  5.14\pm  0.40)+(  3.17\pm  1.16 )\,i$\\
$c^{1/2^+}_{\psi\bar{p} \Lambda,B^-}$         & $  -9.84\pm  2.26                       $ &$  -8.51\pm  2.46                       $ &$  -2.98\pm  2.17                       $\\
$h_{\Xi_c\bar{D},\Xi_c\bar{D}}$               & $ -5.29\pm  0.36$ &$ -4.02\pm  0.30$ & 0 (fixed)\\
$h_{\Lambda_c\bar{D}_s,\Lambda_c\bar{D}_s}$   & $ -3.38\pm  0.22$ & 0 (fixed)        & 0 (fixed)\\
$h_{\Xi_c\bar{D},\Lambda_c\bar{D}_s}$         & $  2.68\pm  0.31$ &$ -1.43\pm  0.95$ & 0 (fixed)\\
$h_{\bar{\Lambda}_c D,\bar{\Lambda}_c D}$     & $  3.30\pm  1.57$ &$  2.81\pm  1.60$ & 0 (fixed)\\
$\Lambda$ (MeV) &1000 (fixed)&1000 (fixed)&1000 (fixed)\\
\end{tabular}
\end{table*}

\begin{figure}[t]
\begin{center}
\includegraphics[width=.5\textwidth]{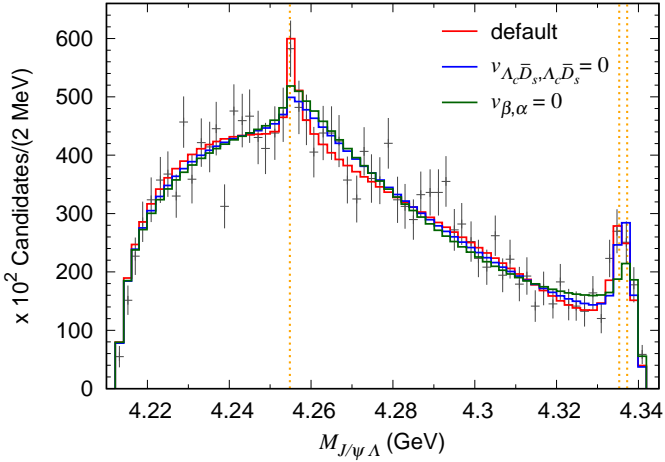}
\end{center}
 \caption{
Comparison of the default model with alternative ones where
 $v_{\Lambda_c\bar{D}_s,\Lambda_c\bar{D}_s}=0$ or
$v_{\beta,\alpha}=0$ for all $\alpha, \beta$ in
 Eq.~(\ref{eq:cont-ptl}). 
Other features are the same as Fig.~\ref{fig:comp-data}(a).
 }
\label{fig:comp-models}
\end{figure}
As the LHCb analysis implies,
the fluctuation at the $\Lambda_c\bar{D}_s$ threshold may be just
statistical and 
$v_{\Lambda_c\bar{D}_s,\Lambda_c\bar{D}_s}$ might be weak.
We thus consider an alternative model by removing
$v_{\Lambda_c\bar{D}_s,\Lambda_c\bar{D}_s}$ from the default model,
refit the data, and show its $M_{J/\psi \Lambda}$ distribution in
Fig.~\ref{fig:comp-models} [blue].
The fit quality is $\chi^2/{\rm ndf}=(58+84+94+30)/(235-8)\simeq 1.19$, similar to the default fit.
An ordinary $\Lambda_c\bar{D}_s$ threshold cusp without a nearby pole is seen.
The default and alternative models have $P_{\psi s}^\Lambda(4338)$
poles on different sheets; see Table~\ref{tab:pole}.
This suggests that higher statistics data should clarify whether a sharp peak exists at the $\Lambda_c\bar{D}_s$ threshold not only for probing $P_{\psi s}^\Lambda(4255)$ but also for constraining the $P_{\psi s}^\Lambda(4338)$ pole structure.

We also consider a case where the $\Xi_c\bar{D}$ interaction has an energy-dependence by 
replacing $h_{\Xi_c\bar{D},\Xi_c\bar{D}}$ in Eq.~(\ref{eq:cont-ptl}) with~\cite{mdu22}
\begin{eqnarray}
h_{\Xi_c\bar{D},\Xi_c\bar{D}}
+
h^\prime_{\Xi_c\bar{D},\Xi_c\bar{D}}
{M^2_{J/\psi \Lambda}-(m_{\Xi_c}+m_{\bar{D}})^2
\over 2(m_{\Xi_c}+m_{\bar{D}})} , 
\label{eq:energy-dep}
\end{eqnarray}
and $h_{\Lambda_c\bar{D}_s,\Lambda_c\bar{D}_s}=0$.
A comparable fit is obtained: $\chi^2/{\rm ndf}=(54+81+95+28)/(235-9)\simeq 1.15$.
This model generates three relevant
$P_{\psi s}^\Lambda(4338)$ poles (Table~\ref{tab:pole}).
A relatively large width of the resonance ($uuu$) 
points to the importance of analyzing
$\Xi^-_b \to J/\psi\Lambda K^-$
where 
the lineshape would reflect the
$P_{\psi s}^\Lambda(4338)$ pole positions more directly.
In $B^-\to J/\psi\Lambda\bar{p}$,
the lineshape from $P_{\psi s}^\Lambda(4338)$
is distorted by the shrinking phase-space.
The different $P_{\psi s}^\Lambda(4338)$ poles
would then be discriminated.

\begin{table}
\caption{\label{tab:para4}
Continued from Table~\ref{tab:para3}.
The second column is for 
the alternative model including an energy-dependent 
$\Xi_c\bar{D}$ interaction with the coupling constant
$h_{\Xi_c\bar{D},\Xi_c\bar{D}}'$.
}
\begin{tabular}{lc}
$h_{\psi\Lambda,\alpha}\,c^{1/2^-}_{\Xi_c \bar{D}\bar{p},B^-}$        & $(  3.85\pm  3.38)+( -4.29\pm 10.31 )\,i$\\
$h_{\psi\Lambda,\alpha}\,c^{1/2^-}_{\Lambda_c \bar{D}_s\bar{p},B^-}$  & $                  ( -0.92\pm  0.49 )\,i$\\
$h_{\psi\bar{p},\bar{\Lambda}_c D}\, c^{1/2^+}_{\bar{\Lambda}_c D   \Lambda,B^-}$ & $(-11.54\pm  3.60)+( 10.83\pm  0.64 )\,i$\\
$c^{1/2^+}_{\psi\bar{p} \Lambda,B^-}$         & $ -14.44\pm  4.21                       $\\
$h_{\Xi_c\bar{D},\Xi_c\bar{D}}$               & $ -2.13\pm  1.89$\\
$h_{\Lambda_c\bar{D}_s,\Lambda_c\bar{D}_s}$   & 0 (fixed)        \\
$h_{\Xi_c\bar{D},\Lambda_c\bar{D}_s}$         & $ -4.29\pm  7.51$\\
$h_{\bar{\Lambda}_c D,\bar{\Lambda}_c D}$     & $  3.55\pm  0.31$\\
$h_{\Xi_c\bar{D},\Xi_c\bar{D}}'$ (MeV$^{-1}$) & $ -0.69\pm  2.37$\\
$\Lambda$ (MeV) &1000 (fixed)                          \\
\end{tabular}
\end{table}

We finally examine if the $P_{\psi s}^\Lambda(4338)$ peak is caused merely by 
the $\Xi_c\bar{D}$ threshold cusp.
A non-pole model, $h_{\beta,\alpha}=0$ in Eq.~(\ref{eq:cont-ptl}), is fitted to the data and shown in Fig.~\ref{fig:comp-models} [green].
While the fit quality, $\chi^2/{\rm ndf}=(69+91+94+33)/(235-5)\simeq 1.25$, is not much worse than the above models overall, the fit in the $P_{\psi s}^\Lambda(4338)$ region is visibly worse. 
Thus, a nearby pole that enhances and sharpens the cusp is necessary to fit the
data.
We made a simple estimate of a statistical significance, and found
that the existence of a nearby pole 
is favored by the LHCb data with 2.6$\sigma$ significance;
see the Supplemental Material for details. 

In Ref.~\cite{pcs4338_ts}, the authors proposed that 
the $P_{\psi s}^\Lambda(4338)$ peak might be caused by 
a mechanism involving a triangle singularity; no nearby pole exists.
Within their model, however, 
the quality of fitting the 
$M_{J/\psi\Lambda}$ distribution
data in the $P_{\psi s}^\Lambda(4338)$ peak
region is similar to what our non-pole model does. 
Thus, the above conclusion should also apply to this
triangle-singularity scenario. 
Yet, the non-pole cusp is sizable and, therefore,
the pole should be located in a position where its impact on the lineshape
is considerably blocked by the $\Xi_c\bar{D}$ branch cut.

\section{Summary}
We analyzed the LHCb data on $B^-\to J/\psi\Lambda\bar{p}$ with diagrams in Fig.~\ref{fig:diag}; 
weak $B^-$ decays are followed by coupled-channel scatterings where $P_{\psi s}^\Lambda$ poles can be developed.
Our default model simultaneously fits the $M_{J/\psi\Lambda}$, $M_{J/\psi\bar{p}}$, $M_{\Lambda\bar{p}}$, and $\cos\theta_{K^*}$ distributions; $\chi^2/{\rm ndf}\sim 1.21$.
We found a $P_{\psi s}^\Lambda(4338)$ pole at $( 4338.2\pm 1.4)-( 1.9\pm 0.5 )\,i$~MeV.
This is the first-time pole determination of the first-discovered hidden-charm pentaquark with
strangeness.
While the pole determination is important in its own right, it also sets the primary basis for investigating the nature of $P_{\psi s}^\Lambda(4338)$.
The data disfavors the $P_{\psi s}^\Lambda(4338)$ structure as just a kinematical effect.
Our default model also fits the fluctuating data at the $\Lambda_c\bar{D}_s$ threshold,
giving a virtual $P_{\psi s}^\Lambda(4255)$ pole at $4254.7\pm 0.4$~MeV.
We also considered alternative fits where $P_{\psi s}^\Lambda(4255)$ does not exist or the $\Xi_c\bar{D}$ interaction has an energy-dependence.
We found $P_{\psi s}^\Lambda(4338)$ poles on different Riemann sheets (Table~\ref{tab:pole}).
The future data should discriminate the different solutions.

\begin{acknowledgments}
This work is in part supported by 
National Natural Science Foundation of China (NSFC) under contracts 
U2032103 and 11625523 (S.X.N.) and under Grants No. 12175239 and 12221005 (J.J.W.),
and also by National Key Research and Development Program of China under Contracts
 2020YFA0406400 (S.X.N., J.J.W.).
\end{acknowledgments}

\begin{center}
 {\bf \large Supplemental Material}
\end{center}

{\bf 1. Statistical assessment to
$P_{\psi s}^\Lambda(4338)$
as purely kinematical threshold cusp}\\

As shown in Fig.~5, the non-pole model (green histogram) gives a visibly worse fit
in the $P_{\psi s}^\Lambda(4338)$ peak region compared to the default
model (red) and the alternative model (blue).
In the following,
we make a simple estimate of a statistical significance with which 
the data disfavors the non-pole model~\cite{stat}.

Suppose we have Model~A fitted to a dataset with $\chi^2=\chi^2_A$. 
Then, we add more mechanisms with $n$ fitting parameters to Model~A and
refit the same dataset, obtaining Model~B with $\chi^2=\chi^2_B$.
We now calculate a $P$-value:
\begin{eqnarray}
\label{eq:p-value}
P(\Delta\chi^2,n) = \int_{\Delta\chi^2}^{+\infty} f_{\chi^2}(z,n)
dz\ ,
\end{eqnarray}
with $\Delta\chi^2 = \chi^2_A-\chi^2_B$.
We have used a $\chi^2$ distribution given by 
\begin{eqnarray}
\label{eq:chi2-fn}
f_{\chi^2}(z,n) = {1\over 2^{n/2}\Gamma(n/2)} z^{n/2-1} e^{-z/2}\ ,
\end{eqnarray}
with the $\Gamma$ function: 
\begin{eqnarray}
\label{eq:Gamma-fn}
\Gamma(x) = \int_0^{+\infty} e^{-t} t^{x-1} dt\ .
\end{eqnarray}
The $P$-value and $Z$-value ($\sigma$) are related by 
\begin{eqnarray}
\label{eq:PZ}
P= {1-{\rm erf}(Z/\sqrt{2})\over 2} \ ,
\end{eqnarray}
with the error function:
\begin{eqnarray}
\label{eq:error-fn}
{\rm erf}(\gamma) = {2\over \sqrt{\pi}}\int_0^\gamma e^{-x^2} dx\ .
\end{eqnarray}

We now apply the above procedure to our models, and see the 
significance of the nearby poles.
We assign the non-pole and alternative models to the above Models~A and B,
respectively, and use $\chi_A^2=69$ and $\chi_B^2=58$
from comparing
with the $M_{J/\psi\Lambda}$ distribution data.
We assume that $\Delta\chi^2=11$ is due to 
the $\Xi_c\bar{D}$-$\Lambda_c\bar{D}_s$ coupled-channel scattering 
and the associated poles and,
thus, the additional fitting parameters 
in the alternative model are 
$h_{\Xi_c\bar{D},\Xi_c\bar{D}}$ and
$h_{\Lambda_c\bar{D}_s,\Lambda_c\bar{D}_s}$ ($n=2$).
Then, we find that 
this additional mechanism 
in the alternative model 
is favored by the data with 2.6$\sigma$ significance. \\

{\bf 2. Parameter values in Tables~\ref{tab:para3} and \ref{tab:para4}}




\end{document}